\documentclass[12pt]{iopart}
\usepackage{epsfig}                                                                 
\usepackage{bm}
\usepackage{amssymb}
\usepackage[pdfstartview=FitH,bookmarksopen=true,bookmarksopenlevel=1]{hyperref}    

\let\operatorname\mathrm
\let\dfrac\frac
\let\text\textrm

\begin{document}
\title[The Enskog--Vlasov equation: A kinetic model describing gas, liquid, and solid]%
      {The Enskog--Vlasov equation:\\A kinetic model describing gas, liquid, and solid}
\author{E. S. Benilov}
 \address{Department of Mathematics and Statistics, University of Limerick, V94 T9PX, Ireland}
 \ead{Eugene.Benilov@ul.ie}
\author{M. S. Benilov}
 \address{Departamento de F\'{\i}sica, CCCEE, Universidade da Madeira, Largo do
Munic\'{\i}pio, 9000 Funchal, Portugal; Instituto de Plasmas e Fus\~{a}o Nuclear,
Instituto Superior T\'{e}cnico, Universidade de Lisboa, Portugal}
 \ead{benilov@uma.pt}

\begin{abstract}
The Enskog--Vlasov (EV) equation is a semi-empiric kinetic model describing gas-liquid phase transitions. In the framework of the EV equation, these correspond to an instability with respect to infinitely long perturbations, developing in a gas state when the temperature drops below (or density rises above) a certain threshold. In this paper, we show that the EV equation describes one more instability, with respect to perturbations with a finite wavelength and occurring at a higher density. This instability corresponds to fluid-solid phase transition and the perturbations' wavelength is essentially the characteristic scale of the emerging crystal structure. Thus, even though the EV model does not describe the fundamental physics of the solid state, it can `mimic' it -- and, thus, be used in applications involving both evaporation and solidification of liquids. Our results also predict to which extent a pure fluid can be overcooled before it definitely turns into a solid.
\end{abstract}
\maketitle

\section{Introduction}

The Enskog--Vlasov (EV) kinetic equation comprises the Enskog collision
integral for dense fluids \cite{Enskog22} and a Vlasov term describing the
van-der-Waals force. The first version of the EV equation \cite{Desobrino67}
was based on the original form of the Enskog integral -- which, as shown in
\cite{LebowitzPercusSykes69}, does not comply with the Onsager relations.
\cite{VanbeijerenErnst73a} proposed a modification of the Enskog integral free
from this shortcoming, which was incorporated in the EV model in
\cite{KarkheckStell81,StellKarkheckVanbeijeren83}. \cite{GrmelaGarciacolin80}
showed that an H-theorem holds for the EV equation only subject to a certain
restriction of its coefficients, and \cite{BenilovBenilov18} proposed a
version of the EV equation that satisfies this restriction and conserves
energy as well (all of the previous versions do not).

Note that, in kinetic models, phase transitions correspond to instabilities.
For the original version of the EV equation, the presence of an instability
has been shown in \cite{Grmela71}, and it was interpreted as gas-liquid phase transition.

In the present paper, we report the results of a more detailed study. Using
the EV\ equation that conserves energy and satisfies an H-theorem, we find
\emph{two} instabilities, with respect to infinite- and finite-wavelength
perturbations -- interpreted as gas-liquid and fluid-solid transitions,
respectively. The latter result comes as a surprise, as the EV equation was
conceived as a tool for modeling of fluids only. We show, however, that it
admits periodic solutions capable of `mimicking' the solid phase.

The present paper has the following structure. In section \ref{section 2}, we
introduce the Enskog--Vlasov equation and, in section \ref{section 3}, carry
out the stability analysis of its spatially homogeneous solutions. The general
results are illustrated by applying them to noble gases in sections
\ref{section 4}--\ref{section 5}.

\section{The Enskog--Vlasov model\label{section 2}}

\subsection{The EV equation}

Consider a fluid of hard spheres of diameter $D$, characterized by the
one-particle distribution function $f(\mathbf{r},\mathbf{v},t)$ where
$\mathbf{r}$ is the position vector, $\mathbf{v}$ the velocity, and $t$ the time.

Let the molecules exert on each other a force with a pair-wise potential
$\Phi(r)$, modeling physically the van der Waals interaction of molecules. Let
$\Phi(r)$ be a monotonically growing function of $r$, so that the van der
Waals force is attractive at all distances. Letting also, without loss of
generality, $\Phi\rightarrow0$ as $r\rightarrow\infty$, we can assume that
$\Phi(r)<0$ for all $r$.

As seen later, the main characteristic of $\Phi$ -- one that affects the
fluid's macroscopic properties -- is%
\begin{equation}
E=-\int\Phi(r)\,\mathrm{d}^{3}\mathbf{r}.\label{2.1}%
\end{equation}
Using $E$, $D$, the molecular mass $m$, and the Boltzmann constant $k_{B}$, we
introduce the following nondimensional variables:%
\[
\mathbf{r}_{nd}=\frac{\mathbf{r}}{D},\qquad\mathbf{v}_{nd}=\left(  \frac
{m}{ED^{3}}\right)  ^{1/2}\mathbf{v},\qquad t_{nd}=\left(  \frac{ED}%
{m}\right)  ^{1/2}t,
\]%
\[
f_{nd}=\frac{k_{B}E^{1/2}D^{3/2}}{m^{3/2}}f,\qquad\Phi_{nd}=\frac{\Phi}%
{ED^{3}}.
\]
Note that, due to (\ref{2.1}), the nondimensional potential $\Phi_{nd}$
satisfies (the subscript $_{nd}$ omitted)%
\begin{equation}
\int\Phi(r)\,\mathrm{d}^{3}\mathbf{r}=-1.\label{2.2}%
\end{equation}
In terms of the nondimensional variables, the Enskog--Vlasov equation has the
form ($_{nd}$ omitted)%
\begin{multline}
\fl\frac{\partial f(\mathbf{r,v},t)}{\partial t}+\mathbf{v}\cdot
\mathbf{\nabla}f(\mathbf{r,v},t)+\mathbf{F}(\mathbf{r},t)\cdot\frac{\partial
f(\mathbf{r,v},t)}{\partial\mathbf{v}}\nonumber\\
=\int\int\left[  \eta(\mathbf{r},\mathbf{r}+\mathbf{\bm{\kappa}}%
,t)\,f(\mathbf{r},\mathbf{v}^{\prime},t)\,f(\mathbf{r}+\mathbf{\bm{\kappa}}%
,\mathbf{v}_{1}^{\prime},t)\right.  \nonumber\\
-\left.  \eta(\mathbf{r},\mathbf{r}-\mathbf{\bm{\kappa}},t)\,f(\mathbf{r}%
,\mathbf{v},t)\,f(\mathbf{r}-\mathbf{\bm{\kappa}},\mathbf{v}_{1},t)\right]
\mathbf{g}\cdot\mathbf{\bm{\kappa}}\,\operatorname{H}(\mathbf{g}%
\cdot\mathbf{\bm{\kappa}})\,\mathrm{d}^{2}\mathbf{\bm{\kappa}}\,\mathrm{d}%
^{3}\mathbf{v}_{1},\label{2.3}%
\end{multline}
where $\operatorname{H}$ is the Heaviside function,%
\begin{equation}
\mathbf{F}(\mathbf{r},t)=-\mathbf{\nabla}\int n(\mathbf{r}_{1},t)\,\Phi
(\left\vert \mathbf{r}-\mathbf{r}_{1}\right\vert )\,\mathrm{d}^{3}%
\mathbf{r}_{1}\label{2.4}%
\end{equation}
is the collective van der Waals force,%
\begin{equation}
n(\mathbf{r},t)=\int f(\mathbf{r},\mathbf{v},t)\,\mathrm{d}^{3}\mathbf{v}%
\label{2.5}%
\end{equation}
is the number density, $\mathbf{\bm{\kappa}}$ is a unit vector parameterizing
all possible orientations of a pair of spheres (molecules) at the moment of
collision, and the post-collision velocities $\left(  \mathbf{v}^{\prime
},\mathbf{v}_{1}^{\prime}\right)  $ are related to the pre-collision ones,
$\left(  \mathbf{v},\mathbf{v}_{1}\right)  $, by%
\begin{equation}
\mathbf{v}^{\prime}=\mathbf{v}+\mathbf{\bm{\kappa}}\left(  \mathbf{g}%
\cdot\mathbf{\bm{\kappa}}\right)  ,\qquad\mathbf{v}_{1}^{\prime}%
=\mathbf{v}_{1}-\mathbf{\bm{\kappa}}\left(  \mathbf{g}\cdot
\mathbf{\bm{\kappa}}\right)  \mathbf{,\qquad g}=\mathbf{v}_{1}-\mathbf{v}%
.\label{2.6}%
\end{equation}
The coefficient $\eta(\mathbf{r},\mathbf{r}_{1},t)$ which appears in the
collision integral is, generally, a functional of $n(\mathbf{r},t)$. It
originates from the main assumption of the EV theory that the two-particle
distribution function $f^{(2)}(\mathbf{r},\mathbf{v},\mathbf{r}_{1}%
,\mathbf{v}_{1},t)$ is related to the singlet $f(\mathbf{r},\mathbf{v},t)$ by%
\[
f^{(2)}(\mathbf{r},\mathbf{v},\mathbf{r}_{1},\mathbf{v}_{1},t)=\eta
(\mathbf{r},\mathbf{r}_{1},t)\,f(\mathbf{r},\mathbf{v},t)\,f(\mathbf{r}%
_{1},\mathbf{v}_{1},t).
\]
Given a specific expressions for $\eta$, equations (\ref{2.3})--(\ref{2.6})
fully determine the evolution of $f$.

There are three approaches to choosing $\eta(\mathbf{r},\mathbf{r}_{1},t)$:

\begin{enumerate}
\item In the original Enskog theory \cite{Enskog22}, $\eta$ is a function of
the number density evaluated at the midpoint between the colliding molecules,
i.e., $n(\frac{1}{2}(\mathbf{r}+\mathbf{r}_{1}),t)$. This function is supposed
to be such that the EV model describes the equation of state (EoS) of the
fluid under consideration with the best possible accuracy.

\item The authors of \cite{VanbeijerenErnst73a} derived $\eta$ from a
hypothesis that the n-particle distribution function is represented by a
product of singlet distributions and (sic!) a factor excluding all states
where the hard spheres overlap. This hypothesis does hold at equilibrium, but
should be considered as approximate otherwise. Another difficulty associated
with this approach is that the resulting $\eta$ is defined through a limiting
procedure involving multiple integrals of increasing order, making it
impossible to solve the EV equation numerically.

\item The authors of \cite{BenilovBenilov18} assumed%
\begin{multline}
\fl\eta(\mathbf{r},\mathbf{r}_{1},t)=1+\sum_{l=2}^{L}c_{l}\int^{l}\left[
{\displaystyle\prod_{i=2}^{l}}
n(\mathbf{r}_{i},t)\,\operatorname{H}(1-\left\vert \mathbf{r}%
-\mathbf{\mathbf{r}}_{i}\right\vert )\,\operatorname{H}(1-\left\vert
\mathbf{r}_{1}-\mathbf{\mathbf{r}}_{i}\right\vert )\right] \nonumber\\
\times\left[
{\displaystyle\prod_{i=2}^{l-1}}
\,\,%
{\displaystyle\prod_{j=i+1}^{l}}
\operatorname{H}(1-\left\vert \mathbf{r}_{i}-\mathbf{\mathbf{r}}%
_{j}\right\vert )\right]
{\displaystyle\prod_{i=1}^{l}}
\mathrm{d}^{3}\mathbf{r}_{i}, \label{2.7}%
\end{multline}
where $\int^{l}$ denotes $l$ repeated integrals, and the coefficients $c_{2}$,
$c_{3}$, $c_{4}$...$c_{L}$ are to be chosen to fit the properties of the fluid
under consideration. Note that the `proper' hard-sphere $\eta$ derived in
\cite{VanbeijerenErnst73a} is a particular case of (\ref{2.7}) -- one with
$L=\infty$ and certain values of $c_{l}$ (which are not easy to calculate).
\end{enumerate}

It turns out that the choice of $\eta$ affects the fundamental properties of
the EV equation.

Consider, for example, the entropy of the system, which is traditionally
assumed \cite{Desobrino67,GrmelaGarciacolin80,GrmelaGarciacolin80b,Grmela81}
to have the form%
\[
S=-\int\int f(\mathbf{r,v},t)\ln f(\mathbf{r,v},t)\,\mathrm{d}^{3}%
\mathbf{v}\,\mathrm{d}^{3}\mathbf{r}+Q[n],
\]
where the non-ideal contribution $Q[n]$ is a functional depending on
$n(\mathbf{r},t)$\footnote{The fact that $Q$ depends only on $n$ and not on
$f$ reflects the hard-sphere nature of the EV model.}. Then, the H-theorem
holds if and only if $Q[n]$ and $\eta$ are inter-related by%
\begin{equation}
\mathbf{\nabla}\frac{\delta Q[n]}{\delta n(\mathbf{r},t)}=-\int\eta
(\mathbf{r},\mathbf{r}_{1},t)\,n(\mathbf{r}_{1},t)\,(\mathbf{r}_{1}%
-\mathbf{\mathbf{r}})\,\delta(\left\vert \mathbf{r}-\mathbf{\mathbf{r}}%
_{1}\right\vert -1)\,\mathrm{d}^{3}\mathbf{r}_{1}\label{2.8}%
\end{equation}
(see \cite{GrmelaGarciacolin80} and, for more detail, Appendix A of
\cite{BenilovBenilov18}). The question of existence of $Q[n]$ as a solution of
equation (\ref{2.8}) for a given $\eta$ is not trivial. If, for example,
$\eta$ is a function of $n\left(  \frac{1}{2}(\mathbf{r+r}_{1}),t\right)  $ --
as in the original Enskog's theory -- (\ref{2.8}) does not seem to have a
solution ofr $Q$. For the versions of $\eta$ suggested in
\cite{VanbeijerenErnst73a,BenilovBenilov18}, on the other hand, it does. In
the latter case, an explicit expression for $Q$ can be found,%
\begin{multline}
\fl Q[n]=-\frac{1}{2}\int\int n(\mathbf{r})\,n(\mathbf{r}_{1}%
)\,\operatorname{H}(1-\left\vert \mathbf{r}-\mathbf{\mathbf{r}}_{1}\right\vert
)\,\mathrm{d}^{3}\mathbf{r}\,\mathrm{d}^{3}\mathbf{r}_{1}\nonumber\\
-\sum_{l=2}^{L}\frac{c_{l}}{l\left(  l+1\right)  }\int^{l}\int n(\mathbf{r}%
)\left[
{\displaystyle\prod_{i=1}^{l}}
n(\mathbf{r}_{i})\,\operatorname{H}(1-\left\vert \mathbf{r}-\mathbf{\mathbf{r}%
}_{i}\right\vert )\right]  \nonumber\\
\times\left[
{\displaystyle\prod_{i=1}^{l-1}}
\,\,%
{\displaystyle\prod_{j=i+1}^{l}}
\operatorname{H}(1-\left\vert \mathbf{r}_{i}-\mathbf{\mathbf{r}}%
_{j}\right\vert )\right]  \mathrm{d}^{3}\mathbf{r}\,%
{\displaystyle\prod_{i=1}^{l}}
\mathrm{d}^{3}\mathbf{r}_{i},\label{2.9}%
\end{multline}
where the coefficients $c_{l}$ are the same as in expression (\ref{2.7}) for
$\eta$.

In this paper, we shall use $\eta$ and $Q$ given by (\ref{2.7}) and
(\ref{2.9}), respectively.

We shall also need the function $\Theta(n)$ related to the functional $Q[n]$
by%
\[
\Theta(n)=-\frac{1}{n}\left(  Q[n]\right)  _{n=\operatorname{const}},
\]
so that (\ref{2.9}) yields%
\begin{equation}
\Theta(n)=\frac{2\pi}{3}n+\sum_{l=2}^{L}\frac{c_{l}A_{l}}{l\left(  l+1\right)
}n^{l}, \label{2.10}%
\end{equation}
where%
\begin{equation}
A_{l}=\int^{l}\left[
{\displaystyle\prod_{i=1}^{l}}
\operatorname{H}(1-\left\vert \mathbf{\mathbf{r}}_{i}\right\vert )\right]
\left[
{\displaystyle\prod_{i=1}^{l-1}}
\,\,%
{\displaystyle\prod_{j=i+1}^{l}}
\operatorname{H}(1-\left\vert \mathbf{r}_{i}-\mathbf{\mathbf{r}}%
_{j}\right\vert )\right]
{\displaystyle\prod_{i=1}^{l}}
\mathrm{d}^{3}\mathbf{r}_{i} \label{2.11}%
\end{equation}
are numeric constants.

$\Theta(n)$ plays an important role in the thermodynamics of EV fluids: in
particular, their EoS is \cite{BenilovBenilov18}%
\begin{equation}
p=nT\left[  1+n\Theta^{\prime}(n)\right]  -\dfrac{1}{2}n^{2}. \label{2.12}%
\end{equation}
where $\Theta^{\prime}=\mathrm{d}\Theta/\mathrm{d}n$.

\subsection{Steady solutions of the EV equation}

Physically, steady (time independent) solutions of the EV equation must have
spatially uniform temperature and zero fluxes of mass, momentum, and energy --
which means that they must be equilibrium states.

To find these, observe that the scattering cross-section in the Enskog
integral does not depend on $\mathbf{v}$ -- as a result, the EV equation is
consistent with the following ansatz:%
\[
f(\mathbf{r},\mathbf{v},t)=\frac{n(\mathbf{r})}{\left(  2\pi T\right)  ^{3/2}%
}\exp\left(  -\frac{\left\vert \mathbf{v}\right\vert ^{2}}{2T}\right)  ,
\]
where $T$ is the temperature. Substituting this ansatz into the EV equation
and carrying out straightforward algebra (see \cite{Grmela71}), we obtain the
following equation for $n(\mathbf{r})$:%
\begin{multline}
\fl\mathbf{\nabla}\left[  \ln n(\mathbf{r})+\frac{1}{T}\int n(\mathbf{r}%
_{1})\,\Phi(\left\vert \mathbf{r}-\mathbf{r}_{1}\right\vert \mathbf{)}%
\,\mathrm{d}^{3}\mathbf{r}_{1}\right]  \\
+\int\eta(\mathbf{r},\mathbf{r}_{1})\,n(\mathbf{r}_{1})\,(\mathbf{r}%
_{1}-\mathbf{r})\,\delta(\left\vert \mathbf{r}_{1}-\mathbf{r}\right\vert
-1)\,\mathrm{d}^{3}\mathbf{r}_{1}=0.
\end{multline}
Subject to (\ref{2.8}), this equation can be integrated,%
\begin{equation}
\ln n(\mathbf{r})+\frac{1}{T}\int n(\mathbf{r}_{1})\,\Phi(\left\vert
\mathbf{r}-\mathbf{r}_{1}\right\vert \mathbf{)}\,\mathrm{d}^{3}\mathbf{r}%
_{1}-\frac{\delta Q[n]}{\delta n(\mathbf{r})}=\mathrm{const}.\label{2.13}%
\end{equation}
This equation coincides with the Euler equation from density functional theory
and also arises in equilibrium statistical mechanics (grand ensemble), where
the term involving $\Phi$ is the functional derivative of the mean field
contribution to the free energy, the $\mathrm{const}$ is the nondimensional
chemical potential divided by $T$, and $Q[n]$ is the excess free energy. The
present derivation shows that $Q[n]$ can also be interpreted as the excess
contribution to, or non-ideal part of, the entropy.

\section{The stability analysis\label{section 3}}

Consider the spatially uniform Maxwellian distribution $f_{M}(\mathbf{v})$. To
examine its stability within the framework of the EV equation, one should let%
\[
f(\mathbf{r},\mathbf{v},t)=f_{M}(\mathbf{v})+\tilde{f}(\mathbf{r}%
,\mathbf{v},t),
\]
where $\tilde{f}(\mathbf{r},\mathbf{v},t)$ is a small perturbation. It is
usually sufficient to examine harmonic perturbations only,%
\begin{equation}
\tilde{f}(\mathbf{r},\mathbf{v},t)=\hat{f}(\mathbf{v})\,\mathrm{e}%
^{ikz+\lambda t}, \label{3.1}%
\end{equation}
where $k$ is the perturbation's wavenumber, $\lambda$ is its growth/decay
rate, and $z$ is one of the spatial coordinates. Substituting (\ref{3.1}) into
the linearized EV equation, one obtains an eigenvalue problem, where $\hat
{f}(\mathbf{v})$ is the eigenfunction and $\lambda$ the eigenvalue. If, for
some $k$, an eigenvalue exists such that $\operatorname{Re}\lambda>0$, the
base state is unstable.

Unfortunately, the outlined procedure implies solving a two-dimensional
integral equation involving the $z$ and normal-to-$z$ components of
$\mathbf{v}$. This equation cannot be solved analytically, and it is even
difficult to be solved numerically.

Instead, we shall only examine \textquotedblleft frozen
waves\textquotedblright, i.e., perturbations with zero growth/decay rate,
$\lambda=0$. They are excellent stability indicators: if a frozen wave with a
wavenumber $k$ exists for a certain state, either a small increase or a small
decrease of $k$ should make it unstable. Thus, the parameter values for which
the first frozen wave bifurcates from the base state corresponds to the onset
of instability.

Admittedly, if $\operatorname{Re}\lambda$ changes sign while
$\operatorname{Im}\lambda\neq0$, this approach fails to detect destabilization
-- but in similar kinetic equations examined for stability so far
\cite{BenilovBenilov16,Fowler19}, this kind of destabilization does not occur.
In the worst-case scenario, one finds some, albeit not all, of the unstable states.

Most importantly, frozen waves in the problem at hand can be found
analytically -- which is incomparably simpler than dealing with the general
perturbations (\ref{3.1}). For the same reason, this kind of stability
analysis is often used in fluid mechanics, in particular, for liquid bridges
(for example, \cite{MeseguerSlobozhaninPerales95,Benilov16}).

\begin{figure}
\begin{flushright}
\includegraphics[width=0.835\textwidth]{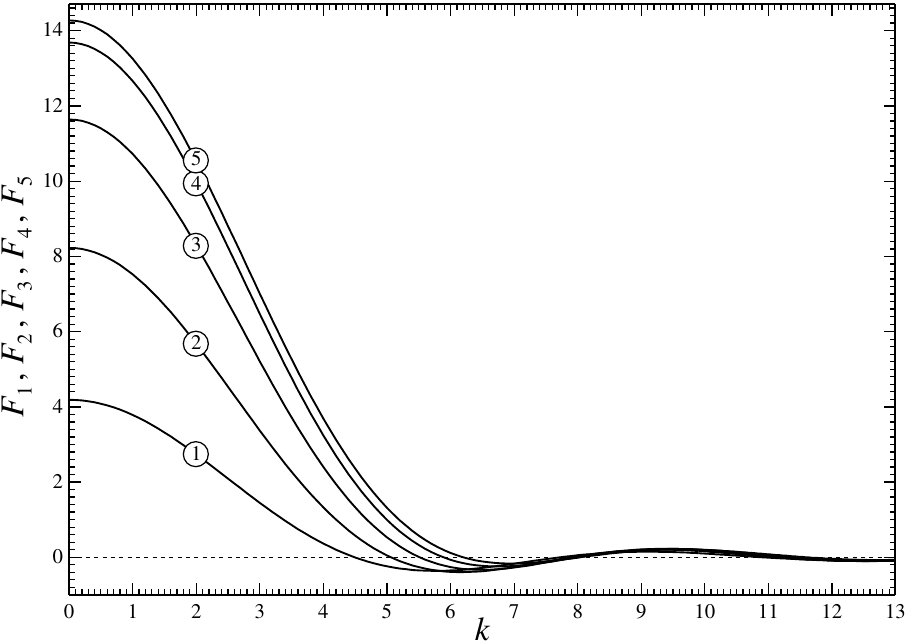}
\end{flushright}
\caption{The functions $F_{l}(k)$ defined by (\ref{3.3}). The curves    are marked with the corresponding value of $l$.}
\label{fig1}
\end{figure}

Since frozen waves are steady, we can search for them using the steady-state
reduction (\ref{2.13}) of the full EV\ equation. To do so, let
\[
n(\mathbf{r})=\bar{n}+\tilde{n}(\mathbf{r}),
\]
where $\bar{n}$ is the density of the base state and $\tilde{n}(\mathbf{r})$
is a perturbation. Substituting expression (\ref{2.10}) for $Q[n]$ into
equation (\ref{2.13}), linearizing it, and letting $\tilde{n}(\mathbf{r}%
)=\mathrm{e}^{ikz}$, we obtain an equation inter-relating $k$, $T$, and
$\bar{n}$ -- which can be written in the form (overbars omitted)%
\begin{equation}
T=-\frac{n\,\hat{\Phi}(k)}{1+nF_{1}(k)+%
{\displaystyle\sum\limits_{l=2}^{L}}
c_{l}n^{l}F_{l}(k)},\label{3.2}%
\end{equation}
where%
\begin{equation}
\fl F_{l}(k)=\int^{l}\left[
{\displaystyle\prod_{j=1}^{l}}
\operatorname{H}(1-\left\vert \mathbf{r}_{j}\right\vert )\right]  \left[
{\displaystyle\prod_{j=1}^{l-1}}
{\displaystyle\prod_{i=j+1}^{l}}
\operatorname{H}(1-\left\vert \mathbf{r}_{j}-\mathbf{r}_{i}\right\vert
)\right]  \cos kz_{l}\,%
{\displaystyle\prod_{j=1}^{l}}
\mathrm{d}^{3}\mathbf{r}_{j},\label{3.3}%
\end{equation}
and%
\[
\hat{\Phi}(k)=%
{\displaystyle\int}
\Phi(r\mathbf{)}\cos kz\,\mathrm{d}^{3}\mathbf{r}.
\]
Note that, due to constraint (\ref{2.2}),%
\begin{equation}
\hat{\Phi}(0)=-1.\label{3.4}%
\end{equation}
Functions $F_{l}(k)$ do not involve any parameters. The first two can be
calculated analytically, and another three have been computed using the
Monte-Carlo method. All five are depicted in figure \ref{fig1}.

Equality (\ref{3.2}) is, essentially, an instability criterion: if a value of
$k$ exists such that (\ref{3.2}) is satisfied for a state $\left(  n,T\right)
$, this state is unstable.

\section{The results\label{section 4}}

In what follows, we shall illustrate criterion (\ref{3.2}) using the values
for the coefficients $c_{l}$, obtained in \cite{BenilovBenilov19} for noble
gases. The series representing $Q$ was truncated at $L=5$, and%
\begin{align}
c_{2} &  =-1.3207,\hspace{1.07cm}c_{3}=9.9308,\label{4.1}\\
c_{4} &  =-18.7526,\qquad c_{5}=13.1406,\label{4.2}%
\end{align}
As seen later, the shape of the Vlasov potential is of little importance, so
we assume, on a more or less ad hoc basis,%
\begin{equation}
\hat{\Phi}(k)=-\frac{1}{1+\left(  Rk\right)  ^{4}},\label{4.3}%
\end{equation}
where $R$ is, physically, the ratio of the spatial scale of the van der Waals
force to the molecule's size. Evidently, expression (\ref{4.3}) complies with
restriction (\ref{3.4}).

\begin{figure}
\begin{flushright}
\includegraphics[width=0.835\textwidth]{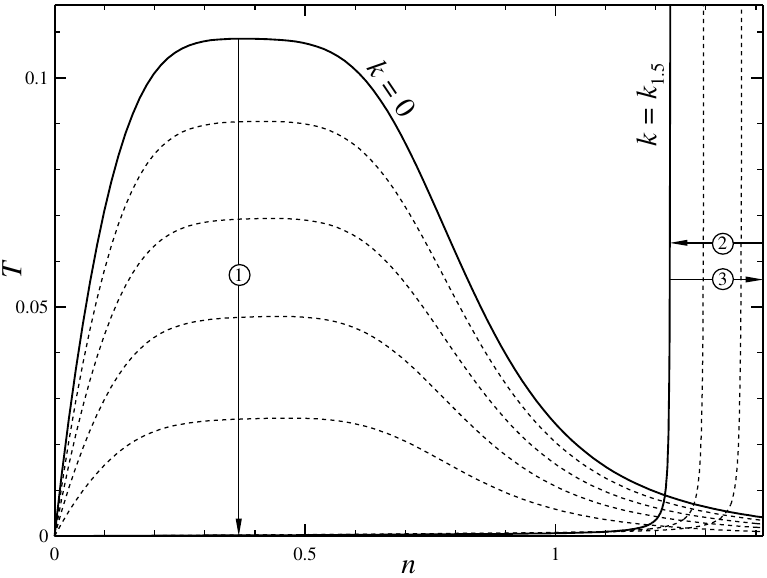}
\end{flushright}
\caption{Existence of frozen waves on the $\left( n,T\right)$ plane. The curves $T(n)$ are determined by (\ref{3.2}), (\ref{4.1})--(\ref{4.3}) with $R=1$. Dotted curves within ranges (1)--(3) correspond to $k$ being within ranges (\ref{4.4})--(\ref{4.6}), respectively. The boundaries of the instability regions are shown by solid lines.}
\label{fig2}
\end{figure}

The stability criterion (\ref{3.2}), (\ref{4.1})--(\ref{4.3}) describes a
one-parameter family of curves $T=T(n)$ with $k$ being the parameter. The
behavior of these curves depends on whether or not the fifth-order polynomial
in $n$ in the denominator of (\ref{3.2}) has positive roots. Computations show
that no more than one such root exists, and it (dis)appear only if
$F_{5}(k)\ $changes sign -- which it does do for infinite sequence of values
of $k$ tending to infinity (see figure \ref{fig1}). Denoting these values by
$k_{1}$, $k_{2}$, $k_{3}$..., we have computed%
\[
k_{1}\approx6.2042,\qquad k_{2}\approx8.0354,\qquad k_{3}\approx11.6014.
\]
A straightforward analysis of expression (\ref{3.2}) shows that, in the range%
\begin{equation}
0<k<k_{1},\label{4.4}%
\end{equation}
the denominator of expression (\ref{3.2}) does \emph{not} have positive roots.
As a result -- and due to quick decay of $\hat{\Phi}(k)$ as $k$ increases --
the curves $T(n)$ `recede' within range (\ref{4.4}) -- see figure \ref{fig2}.
Thus, the curve with $k=0$ determines the boundary of an instability region,
which will be referred to as IR1.

Another instability region (IR2) arises for the range $k_{1}<k<k_{2}$ -- which
can be conveniently subdivided into two subranges,%
\begin{equation}
k_{1}<k<k_{1.5},\label{4.5}%
\end{equation}
with $k_{1.5}\approx7.129$, and%
\begin{equation}
k_{1.5}<k<k_{2}.\label{4.6}%
\end{equation}
As $k$ changes from $k_{1}$ to $k_{1.5}$, the (real positive) root $n_{0}$ of
the denominator of (\ref{3.2}) `travels' from $+\infty$ to $n_{0}\approx
1.230$. Then, when $k$ changes from $k_{1.5}$ to $k_{2}$, $n_{0}$ travels back
to $+\infty$ -- i.e., the boundary of IR2 corresponds to $k=k_{1.5}$. The
corresponding curve $T(n)$ is shown in figure \ref{fig2} together with
examples of curves for $k$ from ranges (\ref{4.5}) and (\ref{4.6}).

A basic analysis of expression (\ref{3.2}) and computations show that the
instability regions corresponding to $\left(  k_{2},k_{3}\right)  $, $\left(
k_{3},k_{4}\right)  $, etc. are all \emph{inside} IR1 and IR2 and, thus, are
physically unimportant.

Finally, if $n\ll1$ (diluted gas), the stability criterion (\ref{3.2}) agrees
with the corresponding results obtained in
\cite{BenilovBenilov16,BenilovBenilov17} for the BGK--Vlasov and
Boltzmann--Vlasov models, respectively.

\section{Discussion\label{section 5}}

For $k=0$ (the boundary of IR1), (\ref{3.2}) and (\ref{3.4}) reduce to%
\[
T=\frac{n}{1+\frac{4\pi}{3}nA_{1}+%
{\displaystyle\sum\limits_{l=2}^{L}}
c_{l}n^{l}A_{l}},
\]
where constants $A_{l}$ are given by (\ref{2.11}). The above expression can be
rewritten in terms of the function $\Theta(n)$ [given by (\ref{2.10})],%
\begin{equation}
T=\frac{n}{1+\left[  n^{2}\Theta^{\prime}(n)\right]  ^{\prime}}. \label{5.1}%
\end{equation}
This representation of the boundary of IR1 turns out to be very useful.

(1) Equation (\ref{5.1}) implies that IR1 does not depend on the specific
shape of the Vlasov potential $\Phi$.

\begin{figure}
\begin{flushright}
\includegraphics[width=0.835\textwidth]{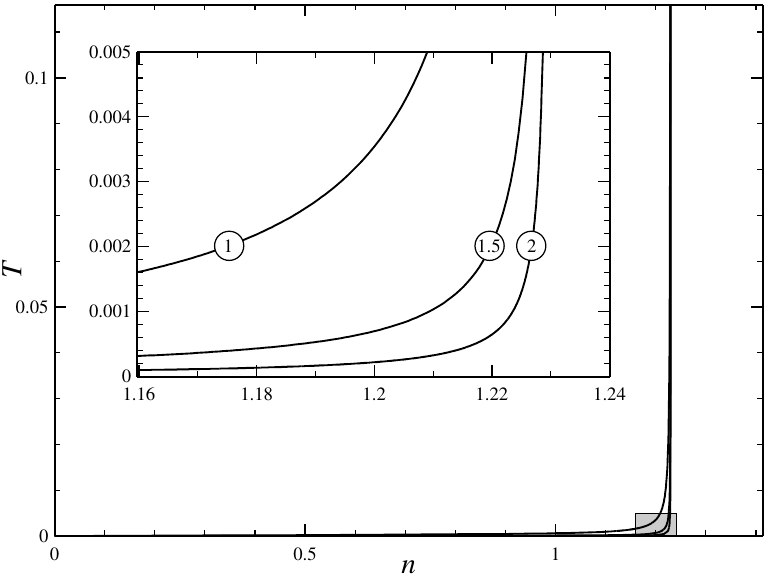}
\end{flushright}
\caption{The dependence of the boundary of IR2 on the parameter $R$ of the Fourier transform (\ref{4.3}) of the Vlasov potential. The inset shows a blow-up of the shaded region of the main panel. The curves are marked with the corresponding values of $R$.}
\label{fig3}
\end{figure}

(2) As for IR2, it does depend on $\Phi$, but this dependence is weak -- which
we illustrate by computing the boundary of IR2 for different values of the
parameter $R$ [which appears in expression (\ref{4.3})] and plotting the
results in figure \ref{fig3}. One can see that, for $R\gtrsim2$, the boundary
of IR2 is virtually indistinguishable from a vertical line. This effect is
even more pronounced if $\hat{\Phi}(k)$ decays exponentially as $k\rightarrow
\infty$.

Given that the van der Waals force is supposed to be long-range (by comparison
with the molecule size), one can assume that $R\gg1$, and thus replace the
boundary of IR2 by a vertical line. Physically, this means that a fluid cannot
be compressed beyond a certain density value no matter what the temperature is.

(3) Using EoS (\ref{2.12}), one can show that the maximum of the function
$T(n)$ given by (\ref{5.1}) corresponds to the critical point.

\begin{figure}
\begin{flushright}
\includegraphics[width=0.835\textwidth]{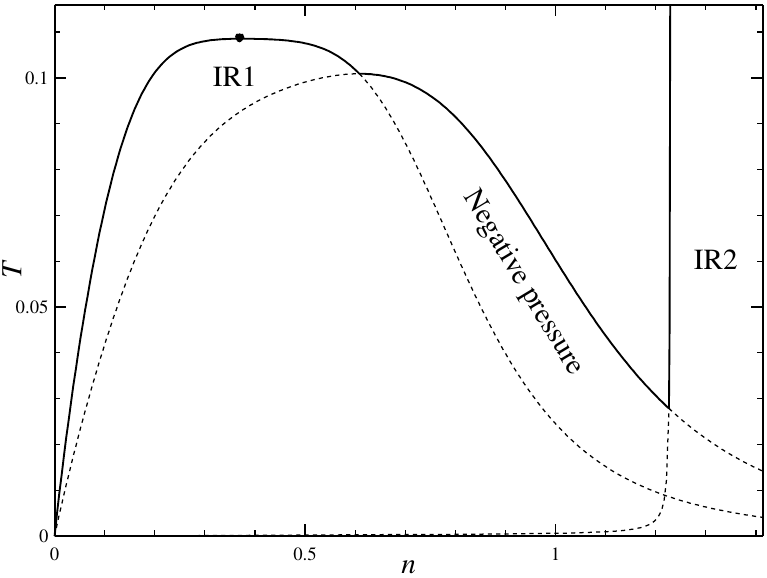}
\end{flushright}
\caption{The stable, physically meaningful fluid states in the $\left(  n,T\right)  $ parameter plane. IR1 and IR2 stand for instability regions 1 and 2, respectively. The black dot marks the critical point.}
\label{fig4}
\end{figure}

(4) Not all of the stable states are physically meaningful, as some of them
correspond to negative pressure. These can be detected using EoS (\ref{2.12}).
For the case (\ref{4.1})--(\ref{4.3}) with $R=1$, the full diagram of stable
and physically meaningful fluid states is shown in figure \ref{fig4}.

(5) As stated in most thermodynamics texts, a non-ideal gas becomes unstable
if
\begin{equation}
\left(  \frac{\partial p}{\partial n}\right)  _{T=\operatorname{const}}<0,
\label{5.2}%
\end{equation}
i.e., if an increase of density gives rise to a decrease of pressure. Applying
this argument to EoS (\ref{2.12}), we recover equation (\ref{5.1}) describing
the boundary of IR1.

IR2, in turn, is located in high-density region -- hence, it may only describe
fluid-solid transitions. Most importantly, the whole boundary of IR2
corresponds to a single value of the perturbation wavenumber, $k_{1.5}$ -- so
that $2\pi/k_{1.5}$ can be identified with the spatial scale of the emerging
crystal. This agrees with the fact that that crystal structure does not depend
on the temperature or density of the fluid state where the transition takes place.

(6) It is well-known that gas-liquid transition typically occurs \emph{before}
criterion (\ref{5.2}) predicts it. The threshold where the actual transition
occurs is determined by the so-called evaporation curve describing the
gas-liquid equilibrium. It is still possible, however, to overcool a gas or
overheat a liquid beyond this threshold, provided they are sufficiently pure.
Thus, the boundaries of the instability regions are essentially the limits to
which one can overcool or overheat a fluid before phase transition occurs.

\begin{figure}
\begin{flushright}
\includegraphics[width=0.835\textwidth]{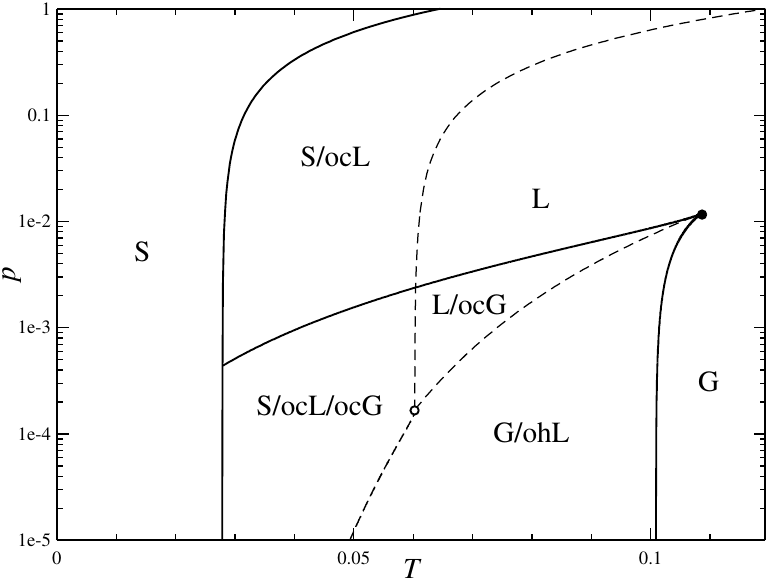}
\end{flushright}
\caption{The phase diagram for argon in the nondimensional $\left( T,p\right)  $ plane. Solid lines correspond to the boundaries of the instability regions computed using the EV model; dashed lines show the empiric evaporation, melting, and sublimation curves \cite{TegelerSpanWagner99}. The critical and triple points are marked by a black dot and small circle, respectively. \textquotedblleft G\textquotedblright, \textquotedblleft L\textquotedblright, and \textquotedblleft S\textquotedblright\ mark the regions where gas, liquid, and solid may exist; the prefixes \textquotedblleft oc\textquotedblright\ and \textquotedblleft oh\textquotedblright\ mean \textquotedblleft overcooled\textquotedblright\ and \textquotedblleft overheated\textquotedblright.}
\label{fig5}
\end{figure}

To illustrate this interpretation, we have redrawn figure \ref{fig4} on the
$\left(  T,p\right)  $ plane, thus turning it into a phase diagram -- see
figure \ref{fig5}. We have also added empirically-derived evaporation,
melting, and sublimation curves (the last two describe the solid-liquid and
solid-gas equilibria, respectively).

The following features of figure \ref{fig5} can be observed:

\begin{itemize}
\item There are two single-phase regions: in the one marked \textquotedblleft
S\textquotedblright, only solid phase exists -- and in the one whose parts are
marked \textquotedblleft L\textquotedblright\ or \textquotedblleft
G\textquotedblright, one of the two fluid phases exists (gas and liquid are
difficult to separate in the latter case, as they can be continuously
transformed one into another).

\item In the transitional zone marked \textquotedblleft
S/ocL\textquotedblright, either solid or overcooled liquid can exist -- and in
the zone \textquotedblleft S/ocL/ocG\textquotedblright, it is either solid or
overcooled liquid, or overcooled gas.

\item In the remaining two zones, \textquotedblleft L/ocG\textquotedblright%
\ and \textquotedblleft G/ohL\textquotedblright, either of the two fluid
phases can exist.
\end{itemize}

\section{Concluding remarks}

In this work, we have used the Enskog--Vlasov model to examine when fluids are
unstable, and with respect to which perturbations. The parameter range of the
instability is illustrated in figure \ref{fig4} on the nondimensional $\left(
n,T\right)  $ plane, and in figure \ref{fig5}, on the $\left(  T,p\right)  $
plane. These figures are the main results of this work.

Note that, in figure \ref{fig5}, we have calculated only the solid curves,
whereas the dashed ones have been obtained by methods of statistical
thermodynamics \cite{TegelerSpanWagner99}. This does not mean that the EV
model cannot be used to calculate the latter: in fact, it \emph{has} been used
for calculating the evaporation curve, producing a result with an error of
only several percent \cite{BenilovBenilov19}. Before calculating the melting
and sublimation curves, however, one should explore periodic solutions of the
EV equation which describe the solid (crystal) state; these solutions
bifurcate from the spatially uniform (fluid) solutions as frozen waves. That
is, we do not claim that the EV model can describe the fundamental physics of
the solid state -- but we do hope that it can `mimic' it given a suitable
choice of the functional $Q[n]$ and the Vlasov potential $\Phi$. In fact, the
Enskog approach to dense fluids has been successfully used for describing
hard-sphere crystals \cite{Kirkpatrick89,KirkpatrickDasErnstPiasecki90} and
studying equilibrium properties of the liquid--solid phase transitions
\cite{RamakrishnanYussouff79,HaymetOxtoby81} (for recent developments in the
latter theory, see
\cite{Archer09,Lutsko12,BaskaranBaskaranLowengrub14,HeinonenAchimKosterlitzYingEtAl16}%
).

Once the EV model is calibrated to deal with all three phases, it would become
an invaluable tool for modeling complex physical problems (e.g., evolution of
liquid films with evaporation and solidification). This is an important point,
as several version of the Enskog--Vlasov kinetic equation have been used for
applications (see
\cite{FrezzottiBarbante17,FrezzottiGibelliLockerbySprittles18} and references therein).

\ack{This work was supported by FCT---Fundação para a Ciência e a Tecnologia of Portugal under Project UID/FIS/50010/2019 and by European Regional Development Fund through the Operational Program of the Autonomous Region of Madeira 2014--2020 under Project PlasMa-M1420-01-0145-FEDER-000016.}\vspace{1cm}

\bibliography{}

\end{document}